\documentclass[preprint,aps,nofootinbib,showpacs,preprintnumbers,amsmath,amssymb]{revtex4}
\usepackage{graphicx,epsfig}
\begin{document}

\preprint{CAVENDISH-HEP-2008-11, DAMTP-2008-72, LPSC 08-125}

\title{NMSSM in disguise: discovering singlino dark matter with soft
leptons at the LHC}

\author{S.\ Kraml}
\affiliation{Laboratoire de Physique Subatomique et de Cosmologie (LPSC),\\
              UJF Grenoble 1, CNRS/IN2P3, 53 Avenue des Martyrs,
              F-38026 Grenoble, France}
\author{A.R.\ Raklev}
\affiliation{DAMTP, Wilberforce Road, Cambridge, CB3 0WA, UK\\
Cavendish Laboratory, JJ Thomson Avenue, Cambridge, CB3 0HE, UK}
\author{M.J.\ White}
\affiliation{Cavendish Laboratory, JJ Thomson Avenue, Cambridge, CB3 0HE, UK\vspace{5mm}}

\date{\today}

\begin{abstract}
We suggest an NMSSM scenario, motivated by dark matter constraints,
that may disguise itself as a much simpler mSUGRA scenario at the
LHC. We show how its non-minimal nature can be revealed, and the
bino--singlino mass difference measured, by looking for soft leptons.
\end{abstract}


\maketitle

\section{INTRODUCTION}

The Next-to-Minimal Supersymmetric Standard Model (NMSSM) provides an
elegant solution to the $\mu$ problem of the MSSM by the addition of a
gauge singlet superfield $\hat S$ \cite{Fayet:1974pd,Nilles:1982dy,
Frere:1983ag,Derendinger:1983bz}. The superpotential of the Higgs
sector then has the form $\lambda\hat S(\hat H_d\cdot\hat
H_u)+\frac{1}{3}\kappa\hat S^3$. When $\hat S$ acquires a vacuum
expectation value, this creates an effective $\mu$ term,
$\mu\equiv\lambda\langle S\rangle$, which is automatically of the
right size, {\it i.e.}\ of the order of the electroweak scale.

The addition of the singlet field leads to a larger particle spectrum
than in the MSSM: in addition to the MSSM fields, the NMSSM contains
two extra neutral (singlet) Higgs fields --\,one scalar and one
pseudo-scalar\,-- as well as an extra neutralino, the singlino. Owing
to these extra states, the phenomenology of the NMSSM can be
significantly different from the MSSM; see Chapter~4 of
\cite{Accomando:2006ga} for a recent review and references. In
particular, the usual LEP limits do not apply to singlet and singlino
states. Moreover, the singlino can be the lightest supersymmetric
particle (LSP) and a cold dark matter candidate.

In this paper, we investigate the LHC signature of an
SPS1a~\cite{Allanach:2002nj}-like scenario supplemented by a singlino
LSP. In such a setup, gluinos and squarks have the `conventional' SUSY
cascade decays into the bino-like neutralino,
${\tilde\chi^0_2}\sim\tilde B$, which then decays into the singlino
LSP, ${\tilde\chi^0_1}\sim\tilde S$, plus a pair of opposite sign
same-flavour (OSSF) leptons. The ${\tilde\chi^0_2}$ decay proceeds
dominantly through an off-shell slepton. A dark matter relic density
of $\Omega h^2\sim 0.1$, compatible with astrophysics measurements, is
obtained if the ${\tilde\chi^0_1}$ and/or ${\tilde\chi^0_2}$
annihilate through pseudo-scalar exchange in the s-channel.

One peculiar feature of this scenario, taking into account
experimental constraints in particular from LEP, is that the mass
difference between ${\tilde\chi^0_1}$ and ${\tilde\chi^0_2}$ turns out
to be small; it reaches at most $\sim 12$ GeV, and is often much
smaller. The leptons originating from the bino decay to the singlino,
\begin{equation}
{\tilde\chi^0_2}\to {\tilde\chi^0_1}\,l^+l^-,
\label{eq:singlino}
\end{equation}
hence tend to be soft. With the recent interest in soft leptons at the
Tevatron~\cite{Aaltonen:2008qz}, searches for these should have appeal
also beyond the model presented here.

In the standard SUSY analysis for LHC events, often requiring
$p_T(l^\pm)>20$~GeV, there is a risk of missing these leptons and
wrongly concluding to have found the MSSM instead of the NMSSM, with
${\tilde\chi^0_2}$ as the erroneous LSP and dark matter
candidate. Discovery of the additional Higgs states will also be very
difficult at the LHC in this scenario.\footnote{The lightest scalar
$S_1$ and pseudo-scalar $A_1$ are mostly singlet states with masses of
about 80--100 GeV and decaying dominantly into $b\bar b$; decays of
heavier Higgs states into them occur only with branching ratios at the
permille level. For a discussion of search strategies, see
\cite{Djouadi:2008uw}.}  The aim of this paper is to explore the
possibility of detecting the ${\tilde\chi^0_2}\to
{\tilde\chi^0_1}l^+l^-$ decay, and measuring the bino--singlino mass
difference, by looking for soft di-leptons at the LHC.  Some
preliminary work on this scenario was carried out in~\cite{:2008gv}.

In Section~\ref{sect:scenario} we begin by describing the particular
NMSSM scenario under investigation, and define five benchmark points
typifying the small bino--singlino mass difference. We go on to
discuss the Monte Carlo simulation of these benchmark points in
Section~\ref{sect:mc}, using the fast simulation of a generic LHC
detector as a basis for studying the possibility of detecting the
resulting soft leptons in LHC collisions. In Section~\ref{sect:mass}
we further discuss the extraction of mass constraints on the singlino
from the di-lepton invariant mass distribution using shape fitting,
before we conclude in Section~\ref{sect:conclusions}.

\section{The NMSSM scenario}
\label{sect:scenario}

We use the {\tt NMHDECAY}~\cite{Ellwanger:2004xm,Ellwanger:2005dv}
program to compute the NMSSM mass spectrum and Higgs bran\-ching ra\-tios,
and to evaluate the LEP bounds; {\tt SPHENO}~\cite{Porod:2003um} is
used to calculate the sparticle branching ratios, and 
{\tt MICROMEGAS}~\cite{Belanger:2005kh,Belanger:2006is} 
for the relic density.
The SUSY-breaking parameters of our scenario are listed in
Table~\ref{tab:MSSMparameters}. The main difference from the familiar
mSUGRA scenario SPS1a~\cite{Allanach:2002nj} is that we choose a
larger $M_1=0.5M_2=120$~GeV, leading to bino and wino masses of
$115$~GeV and $222$~GeV, respectively, in order to evade LEP bounds
when adding the singlino and singlet Higgses. In the original SPS1a
scenario, the bino and wino masses are $96$ and $177$~GeV. The
resulting SUSY spectrum, with the exception of the NMSSM-specific
masses, is shown in Table~\ref{tab:MSSMmasses}.

\begin{table}[t]
\begin{center}
\begin{tabular}{lccccccccccccc}
\hline
Parameter & $M_1$ & $M_2$ & $M_3$ & $\mu_{\rm eff}$ 
                  & $M_{\tilde L_{1,3}}$ & $M_{\tilde E_1}$ & $M_{\tilde E_3}$
                  & $M_{\tilde Q_1}$ & $M_{\tilde U_1}$ & $M_{\tilde D_1}$ 
                  & $M_{\tilde Q_3}$ & $M_{\tilde U_3}$ & $M_{\tilde D_3}$\\    
Value [GeV] & ~120~ & ~240~ & ~720~ & ~360~ & ~195~ 
                    & ~136~ & ~133~ & ~544~ & ~526~ & ~524~ & ~496~ & ~420~ & ~521~\phantom{.}\\
\hline
\end{tabular}
\caption{Input parameters in for our SPS1a-like scenario.  
   The NMSSM-specif\/ic parameters $\lambda$, $\kappa$, $A_\lambda$ 
   and $A_\kappa$ are given in Table~\ref{tab:NMSSMpoints}.}
\label{tab:MSSMparameters}
\end{center}
\end{table}

\begin{table}[tb]
\begin{center}
\begin{tabular}{lccccccccccc}
\hline
Particle   & $\tilde\chi^0_2$ & $\tilde\tau_1$ & $\tilde e_R$ & $\tilde e_L$
           & $\tilde\tau_2$ & $\tilde\chi^0_3$ & $\tilde\chi^0_4$
           & $\tilde\chi^0_5$ & $\tilde t_1$ & $\tilde q_{L,R}$
           & $\tilde g$ \\
Mass [GeV] & ~115~ & ~132~ & ~143~ & ~201~ & ~205~ & ~222~ & ~365~ 
                    & ~390~ & ~397~ & ~550--570~ & ~721~\phantom{.} \\
\hline
\end{tabular}
\caption{Mass spectrum of our SPS1a-like scenario;
$m_{\tilde\mu_{L,R}}=m_{\tilde e_{L,R}}$ and
$m_{\tilde\chi^\pm_{1,2}}\simeq m_{\tilde\chi^0_{3,5}}$. The LSP and
Higgs masses depend on the NMSSM-specific parameters and are given in
Table~\ref{tab:NMSSMpoints}.}
\label{tab:MSSMmasses}
\end{center}
\end{table}

To obtain a singlino LSP, we further choose $\lambda \sim 10^{-2}$ and
$\kappa\sim 0.1\lambda$. This way $\tilde\chi^0_1\sim 99\%\,\tilde S$,
and $m_{\tilde\chi^0_2}$ hardly varies with $\lambda$ and $\kappa$
($\Delta m_{\tilde\chi_2^0}\sim 0.1$~GeV). In addition, the trilinear
Higgs couplings $A_\lambda$ and $A_\kappa$ are chosen such that
$m_{\tilde\chi^0_i}+m_{\tilde\chi^0_j}\sim m_{A_2}$ for at least one
combination of $i,j=1,2$, in order to achieve a dark matter density of
$0.094\le\Omega h^2\le 0.135$~\cite{Hamann:2006pf}. We thus obtain a
set of NMSSM parameter points with varying $\Delta m \equiv
m_{\tilde\chi^0_2}-m_{\tilde\chi^0_1}$.

\begin{table}[tb]
\begin{center}
\begin{tabular}{cccccccccccc}
\hline
  Point & ~$\lambda\,[10^{-2}]$~ & ~$\kappa\,[10^{-3}]$~ & $A_\lambda$ & $A_\kappa$ 
           & $m_{\tilde\chi^0_1}$  & $m_{A_1}$ & $m_{A_2}$ & $m_{S_1}$   
           & $\Omega h^2$ &  $\Gamma({\tilde\chi^0_2})$\\
\hline
 A & $1.49$ & $2.19$ & ~$-37.4$~ & ~$-49.0$~ & ~$105.4$~ & ~$88$~ & ~$239$~ & ~$89$~  
     & ~$0.101$~  & ~$7\times10^{-11}$~ \\ 
 B & $1.12$ & $1.75$ & ~$-42.4$~ & ~$-33.6$~ & ~$112.1$~ & ~$75$~ & ~$226$~ & ~$100$~  
     & ~$0.094$~  & ~$9\times10^{-13}$~ \\ 
 C & $1.20$ & $1.90$ & ~$-39.2$~ & ~$-53.1$~ & ~$113.8$~ & ~$95$~ & ~$256$~ & ~$97$~  
    & ~$0.094$~  & ~$1\times10^{-13}$~ \\ 
 D & $1.47$ & $2.34$ & ~$-39.2$~ & ~$-68.9$~ & ~$114.5$~ & ~$109$ & ~$259$~ & ~$92$~  
    & ~$0.112$~  & ~$4\times10^{-14}$~ \\ 
 E & $1.22$ & $1.95$ & ~$-44.8$~ & ~$-59.1$~ & ~$114.8$~ & ~$101$~ & ~$219$~ & ~$96$~  
    & ~$0.096$~  & ~$8\times10^{-15}$~ \\ 
\hline 
\end{tabular}
\caption{NMSSM benchmark points used in this study. Masses and other
dimensionful quantities are in [GeV]. The other parameters/the rest of
the spectrum are/is given in the previous tables.}
\label{tab:NMSSMpoints}
\end{center}
\end{table}

The five points used in this study are summarised in
Table~\ref{tab:NMSSMpoints}. Points A--E have $\Delta m=9.71$, $3.05$,
$1.45$, $0.87$ and $0.60$ GeV, respectively.  The SM-like second neutral
scalar Higgs, $S_2$, has a mass of 115~GeV for all these points,
consistent with the LEP limit of 114.4~GeV~\cite{Barate:2003sz}. 
By contrast, the lightest neutral scalar $S_1$ and the lighter pseudo-scalar 
$A_1$ are mostly singlet states, and can hence be lighter than 114.4~GeV.
Concerning the efficient neutralino annihilation needed to achieve an
acceptable dark matter density, for Point A the dominant channel is
$\tilde\chi^0_2\tilde\chi^0_2\to b\bar b$, contributing $88\%$ to
the thermally averaged annihilation cross section times relative velocity, 
$\langle\sigma v\rangle \propto 1/(\Omega h^2)$.
For Point B, $\tilde\chi^0_1\tilde\chi^0_1$,
$\tilde\chi^0_1\tilde\chi^0_2$ and $\tilde\chi^0_2\tilde\chi^0_2$
annihilation to $b\bar b$ contribute $10\%$, $15\%$, and $50\%$,
respectively.  Point C has again dominantly
$\tilde\chi^0_2\tilde\chi^0_2$, while Point D has about 50\%
$\tilde\chi^0_2\tilde\chi^0_2$ and 35\% $\tilde\chi^0_1\tilde\chi^0_2$
annihilation. Finally, for Point E, $\tilde\chi^0_1\tilde\chi^0_1$,
$\tilde\chi^0_1\tilde\chi^0_2$ and $\tilde\chi^0_2\tilde\chi^0_2$
annihilation to $b\bar b$ contribute $29\%$, $34\%$, and $13\%$,
respectively.

\begin{figure}
\begin{center}
\includegraphics[width=0.5\textwidth]{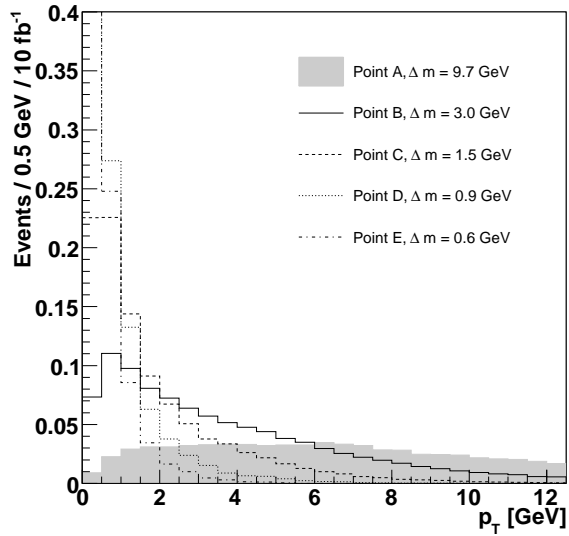}
\caption{$p_T$ distributions for leptons from the decay
$\tilde\chi^0_2\to\tilde\chi^0_1l^+l^-$ in benchmark points A--E. All
distributions are normalised to unity over the whole momentum range.}
\label{fig:truept}
\end{center}
\end{figure}

Assuming that the non-MSSM nature of the Higgs sector will be very
time and integrated luminosity consuming to determine at the LHC, or
even be difficult to clarify at all, early signatures of this scenario
will have to rely on the leptons produced by the two
${\tilde\chi^0_2}\to {\tilde\chi^0_1}l^+l^-$ decays present in the
vast majority of SUSY events. Figure~\ref{fig:truept} shows the $p_T$
distributions of these leptons for all five benchmark
points.\footnote{For details of the Monte Carlo simulation used, see
Section~\ref{sect:mc}.} Clearly, cuts on lepton transverse momentum of
even 10 GeV will remove the wast majority of events for points B--E,
and hence remove the one remaining clue to the non-minimal nature of
the scenario. However, one should also notice that the distributions
have considerable tails beyond the simple mass difference $\Delta m$,
due to the boost of the $\tilde\chi^0_2$. Thus, a reduction in the
lepton $p_T$-cut holds the promise of giving considerable extra reach
in this scenario.

\section{MONTE CARLO ANALYSIS}
\label{sect:mc}

We perform a Monte Carlo simulation of the benchmark points described
above by generating our SUSY signal with {\tt
PYTHIA~6.413}~\cite{Sjostrand:2006za} and SM background events with
{\tt HERWIG~6.510}~\cite{Corcella:2000bw,Corcella:2002jc}, interfaced
to {\tt ALPGEN 2.13}~\cite{Mangano:2002ea} for production of high jet
multiplicities and {\tt JIMMY 4.31}~\cite{Butterworth:1996zw} for
multiple interactions. The generated events are then put through a
fast simulation of a generic LHC detector, {\tt
AcerDET-1.0}~\cite{Richter-Was:2002ch}.

Although {\tt PYTHIA} does not contain a framework for generating
NMSSM events {\it per se}, it has the capability to handle the NMSSM
spectrum and its decays. Since our scenario predicts the same dominant
cross section as in the MSSM, namely gluino and squark
pair-production, with negligible interference from the non-minimal
sector, we use the built-in MSSM machinery for the hard process, and
take the conservative approach of generating only events with squark
and gluino production. For the signal, {\tt PYTHIA} gives a LO cross
section of 24~pb, and 240\,000 events are generated per benchmark
point, corresponding to 10~fb$^{-1}$ of data.

For the SM background we have generated a wide variety of samples that
in addition to two, possibly soft, OSSF isolated leptons at low
invariant mass, could potentially yield the hard jets and missing
energy expected for SUSY events. These consist of $n$-jet QCD samples,
generated with jet-parton matching using {\tt ALPGEN}, and the
production of $W$, $Z$, $WW$, $WZ$, $ZZ$, $b\bar b$, and $t\bar t$
with $n$ additional jets ($n\leq 3$). In addition to this we have also
looked at the Drell-Yan production of lepton pairs at low invariant
masses (3~GeV $<m_{ll}<20$~GeV) with $n$ additional jets.

These samples were passed through the {\tt AcerDET} detector
simulation. For the scenario we consider {\tt AcerDET} gives a
reasonable description of the response of an LHC detector, with the
exception of soft objects. Given the importance of soft leptons to our
study, we therefore modified {\tt AcerDET} as follows:
\begin{enumerate}
\item
the $p_T$ threshold for leptons was lowered to $2$~GeV;
\item
the lepton momentum resolutions used were parameterised from the
results of a full simulation of the ATLAS detector, as presented
in~\cite{atlcomphys2007102};\footnote{For muons we use the results for
combined muon system and inner detector tracks with $|\eta| <
1.1$. The electrons are smeared according to a pseudo-rapidity
dependent parametrisation.}
\item
we applied parameterised lepton reconstruction efficiencies extracted
from results given in~\cite{atlcomphys2007102}.\footnote{Again we use
results from combined muon system and inner detector tracks for the
muons, where muons down to $1$~GeV have been simulated. For electrons
we use the efficiency of so-called ``tight cuts'', defined
in~\cite{atlcomphys2007102}, in busy physics events.}
\end{enumerate}
This simulation then incorporates the most relevant effects for the
analysis, such as a sensible description of the rapidly deteriorating
lepton momentum resolution for electrons at $p_T<20$~GeV. Also, the
applied reconstruction efficiencies fall off steeply for low $p_T$,
and are different for electrons and muons, adding another degree of
realism to the reconstruction of an invariant mass distribution using
both lepton flavors.

However, there are some issues regarding detector performance at low
$p_T$ that are not modelled by these additions, chiefly the
introduction of fake electrons through, e.g., the mis-identification
of charged pions. In particular, one might worry about the potential
of pure QCD events to fake our signal because of the huge cross
section in conjunction with detector effects. To improve on the
parameterisations used here one would need a full simulation of the
detector, or even efficiencies from data. We shall show below that all
backgrounds with pairs of uncorrelated leptons may in principle be
estimated from data, assuming lepton universality or some knowledge of
the degree of non-universality. Therefore, the purity of the
reconstructed electron sample is less important than the efficiency
for reconstructing the sample in the first place.

We carry out our analysis along the lines of the `standard' di-lepton
edge analysis~\cite{Hinchliffe:1996iu,Bachacou:1999zb}, see
also~\cite{Allanach:2000kt,Gjelsten:2004ki}. To isolate the SUSY
signal from SM background we apply the following cuts:
\begin{itemize}
\item
Require at least three jets with $p_T>150,\,100,\,50$~GeV.
\item
Require missing transverse energy $\not\!\! E_T > \max(100~{\rm
GeV},0.2M_{\rm eff})$, where the effective mass $M_{\rm eff}$ is the
sum of the $p_T$ of the three hardest jets plus missing energy.
\item
Require two OSSF leptons with $p_T>20,\,10$~GeV.
\end{itemize}
After these cuts the background is small compared to the number of
SUSY events. For all five benchmark points the resulting di-lepton
invariant mass distributions have the expected edge structure at $\sim
80$~GeV from the decay chain
\begin{equation}
\tilde\chi_3^0\to\tilde l_L^\pm l^\mp\to\tilde\chi_2^0l^+l^-,
\label{eq:slepton}
\end{equation}
and another, much less visible structure, at $\sim 110$~GeV, due to
the same decay through a right handed slepton. These can be treated in
the usual manner to extract two relationships between the four
involved SUSY masses, based on the position of the endpoints.

\begin{figure}
\begin{center}
\includegraphics[width=\textwidth]{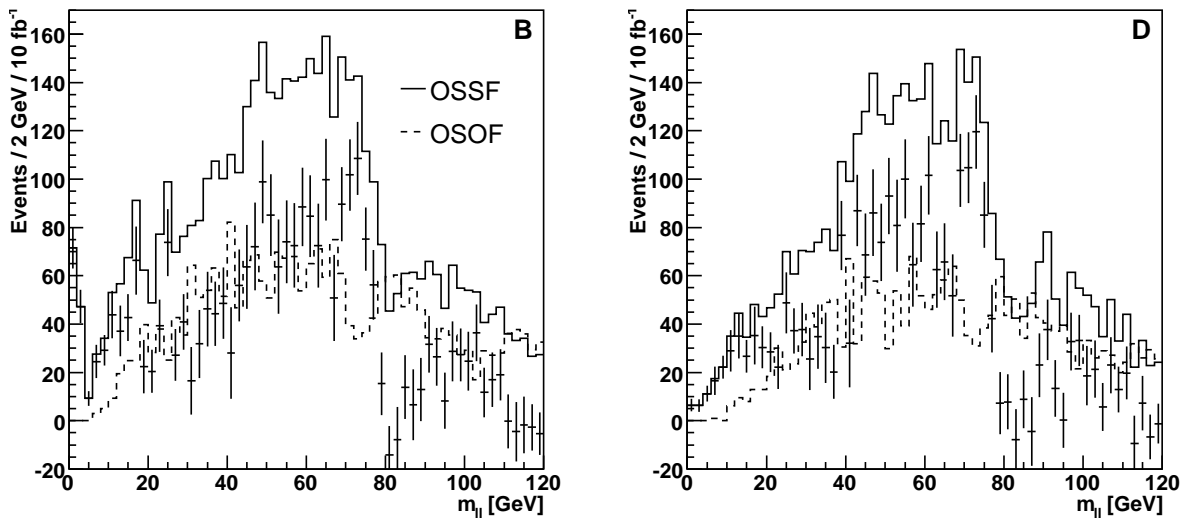}
\caption{Di-lepton invariant mass distributions for point B (left) and
point D (right) with standard lepton $p_T$ cuts. Shown are OSSF
(solid), OSOF (dashed) and subtracted (with error bars)
distributions.}
\label{fig:subtracted_stdlepc}
\end{center}
\end{figure}

For benchmark points A and B there is also an excess of events at low
invariant mass coming from the decays of $\tilde\chi_2^0$ to
singlinos, but for points C, D and E the OSSF lepton selection cut has
removed all trace of the $\tilde\chi_2^0$ decay. This is demonstrated
in Figure~\ref{fig:subtracted_stdlepc}, showing the di-lepton
invariant mass distribution for two of the benchmark points. In
scenarios like C, D and E, one would therefore risk missing the
singlino and taking the $\tilde\chi_2^0$ to be the LSP dark matter
candidate.\footnote{In fact, our SPS1a-like scenario is an optimistic
one for soft leptons under the standard cuts, in that there are extra
leptons at hand from the longer decay chain (\ref{eq:slepton}) to
fulfil the cut requirement.} For a further breakdown of the content of
the di-lepton invariant mass distribution in such scenarios,
see~\cite{:2008gv}.

\begin{figure}
\begin{center}
\includegraphics[width=\textwidth]{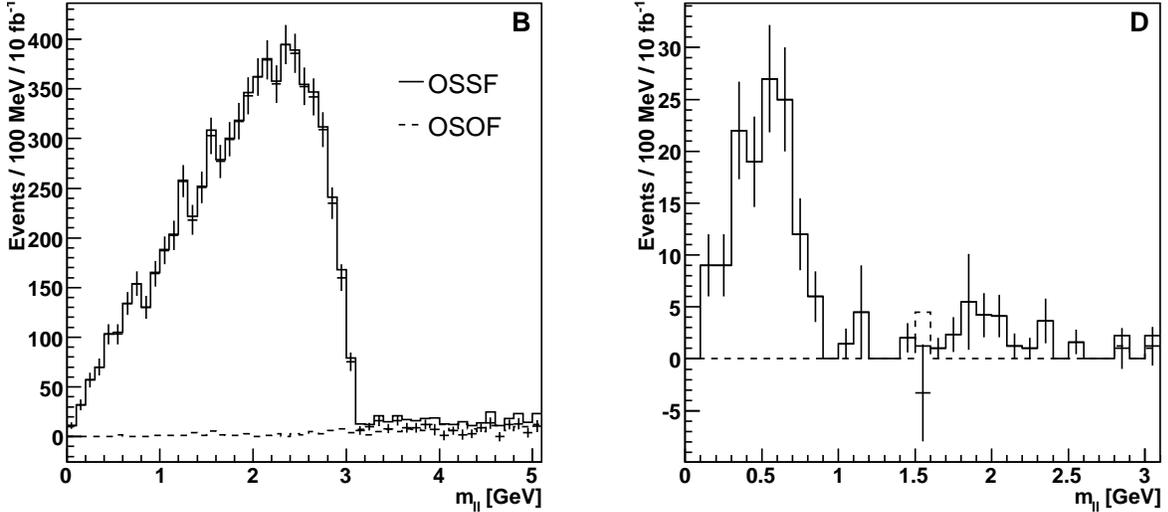}
\caption{Di-lepton invariant mass distributions for point B (left) and
point D (right). Shown are OSSF (solid), OSOF (dashed) and subtracted
(with error bars) distributions.}
\label{fig:subtracted}
\end{center}
\end{figure}

It is clear that to increase sensitivity to the disguised NMSSM
scenario, one needs to lower the lepton $p_T$ cuts. Due to the hard
requirements on jets and missing energy, the vast majority of these
events should still pass detector trigger requirements. However, lower
cuts come with the possibility of large increases in SM
backgrounds. Most of this background, that from uncorrelated leptons,
can in principle be removed by subtracting the corresponding opposite
sign opposite-flavour (OSOF) distribution, assuming lepton
universality. However, larger backgrounds will increase the
statistical error. In addition, a soft lepton sample is more
vulnerable to the introduction of non-universality from e.g.\ pion
decays. The result of lowering the $p_T$ requirement on leptons to
$2$~GeV, after application of the reconstruction efficiency, is shown
in the left and right panels of Fig.~\ref{fig:subtracted}, for
benchmark points B and D respectively. While there is an increase in
backgrounds, the effect on the signal is much more significant. For
both benchmarks, the decay to the singlino is now visible as a large
excess at low invariant masses.

\begin{figure}
\begin{center}
\includegraphics[width=0.5\textwidth]{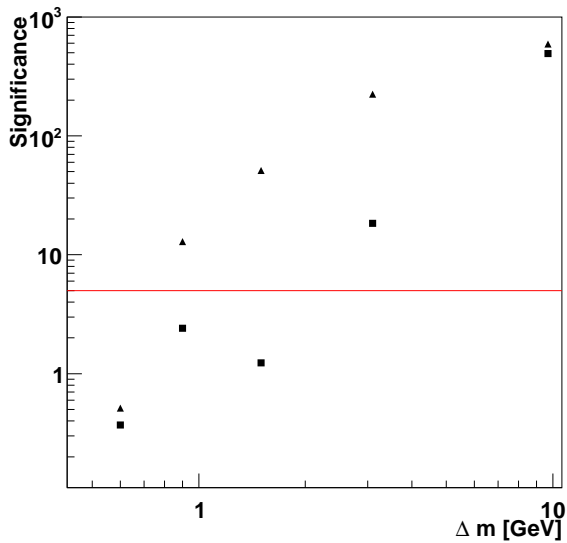}
\caption{Significance vs. bino--singlino mass difference $\Delta m$.
We show the significance both with the standard lepton $p_T$ cut
(squares) and the $2$~GeV lepton cut (triangles).}
\label{fig:reach}
\end{center}
\end{figure}

To quantify the potential for discriminating between the disguised
NMSSM and mSUGRA, we show in Fig.~\ref{fig:reach} the significance
$S/\sqrt{B}$ of any excess at low invariant masses as a function of
$\Delta m$ for all five benchmark points and for both the standard and
$2$~GeV lepton $p_T$ cut. The expected number of events $B$, in the
absence of any signal, is estimated by the OSOF distribution plus the
expected number of mSUGRA low invariant mass leptons from a fit to the
edge at $\sim 80$~GeV, continued down to low invariant masses. This
means that the expected number of events can be determined entirely
from data. The number of signal events $S$ are the events in excess of
this. $S/\sqrt{B}$ is evaluated for low invariant masses, taking
$m_{ll}<10$~GeV as an upper limit. The exact significance will
naturally depend on the interval chosen, but $10$~GeV should in any
case be conservative. At low significance there is, as expected, some
fluctuation in the significance due to the random nature of the signal
generation and background generation. From Fig.~\ref{fig:reach} we
find that we should be able to observe a significant excess down to
$\Delta m\simeq 0.8$~GeV, under the assumptions on lepton efficiencies
described above.\footnote{For such small mass differences we may also
begin to see neutral displaced vertexes, cf.~\cite{Ellwanger:1998vi}.} 
However, it is worth noting that even with the standard lepton cuts
one should be sensitive to mass differences down to $2-3$~GeV.

\section{MASS CONSTRAINTS}
\label{sect:mass}

In the standard di-lepton analysis the edges at $\sim 80$~GeV and
$\sim 100$~GeV are used to determine the relationship between the
neutralino and slepton squared-mass differences, in our scenario
$m_{\tilde\chi^0_3}^2-m_{\tilde l_{L/R}}^2$ and $m_{\tilde
l_{L/R}}^2-m_{\tilde\chi^0_2}^2$, and the slepton squared-mass
$m_{\tilde l_{L/R}}^2$. With the addition of further edges from longer
decay chains, the individual masses $m_{\tilde\chi^0_3}$, $m_{\tilde
l_{L/R}}$ and $m_{\tilde\chi^0_2}$ can be constrained, although mass
differences are determined much more precisely. For the SPS1a
benchmark point, with similar masses to our scenario, one finds that a
precision of $\sim 4\%$ is achievable on the masses of the neutralinos
and sleptons involved, when the measurement is systematics
dominated~\cite{Gjelsten:2004ki}.

In the same manner we could also attempt to extract information on the
singlino by determining the position of the edge at low invariant
masses, giving access to the mass difference
$m_{ll}^{\max}=m_{\tilde\chi^0_2}-m_{\tilde\chi^0_1}$, for the
three-body decay. Since the shape of an invariant mass distribution
{\it a priori} contains more information than an endpoint, it could be
hoped that a fit to the whole distribution would further constrain the
SUSY parameters involved, e.g. setting the scale of the masses as well
as their
difference~\cite{Miller:2005zp,Gjelsten:2006as,Gjelsten:2006tg}. In
fact, the full matrix element for the $\tilde\chi^0_2$ three-body
decay via a virtual slepton, as calculated in~\cite{Bartl:1986hp}, is
used in {\tt PYTHIA}. From Eq.~(11) of~\cite{Bartl:1986hp} we can see
that the invariant mass distribution, in addition to the neutralino
masses, also depends on the left and right handed slepton masses and
their widths.

\begin{figure}
\begin{center}
\includegraphics[width=\textwidth]{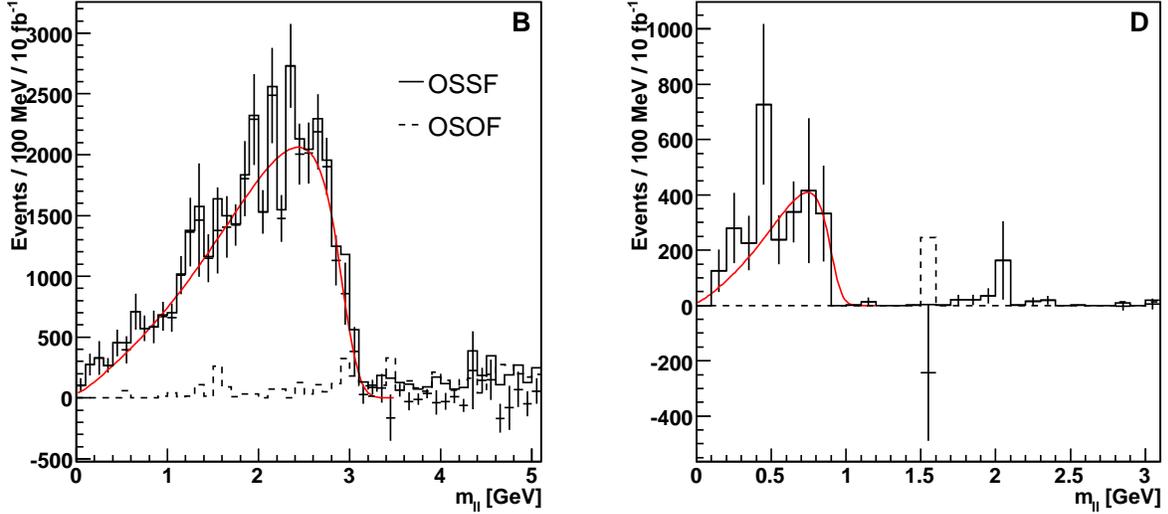}
\caption{Di-lepton invariant mass distributions for point B (left) and
point D (right) after re-weighting with lepton efficiencies. Fits (in
red) are described in the text.}
\label{fig:subtracted_weighted}
\end{center}
\end{figure}

We perform fits to the di-lepton invariant mass distributions at low
invariant masses with a Gaussian smearing of the shape given
in~\cite{Bartl:1986hp}, under the assumption that effectively only one
slepton contributes.\footnote{For our benchmark points the right
handed slepton contributes $\sim 90$\% to the decay amplitude. We have
checked that the fits are completely insensitive to whether there are
one or two sleptons participating.} The Gaussian smearing is meant to
emulate smearing by finite detector resolution. The results for
benchmark points B and D are shown in
Fig.~\ref{fig:subtracted_weighted}. In subtracting the OSOF
distribution before fitting, we have taken into account the effective
lepton non-universality induced by the difference in electron and muon
efficiencies. This is done by re-weighting pairs of leptons with the
inverse of their combined efficiencies, according to the lepton
momenta involved. This effectively unfolds the non-universality effects
on the invariant mass distribution from the differing efficiencies, at
the cost of increasing statistical errors due to large weights. It
should be a simple extension to include geometry dependent
efficiencies into this re-weighting. Naturally, the re-weighting can
only be effective if the errors on the measured lepton efficiencies
are small compared to the other errors involved in the fit. The
resulting differences in shape can clearly be seen by comparing
Figs.~\ref{fig:subtracted} and
\ref{fig:subtracted_weighted}.

Despite our hopes, we find that the shape fits do not constrain the
slepton width at all, nor do they constrain the absolute mass scale
significantly. For benchmark point A, with the largest statistics, it
indicates a singlino mass of $m_{\tilde\chi_1^0}=83.2\pm
44.1$~GeV. However, when parametrised in terms of the bino--singlino
and slepton--singlino mass differences, keeping also the scale as free
parameter, the fits give quite good bounds, which can be found in
Table~\ref{tab:masses}.

\begin{table}[tb]
\begin{center}
\begin{tabular}{lcccc}
\hline
Benchmark point
                       & A              & B              & C              & D         \\ \hline
$m_{\tilde\chi^0_2}-m_{\tilde\chi^0_1}$
                       & ~$9.77\pm 0.03$~ & ~$2.98\pm 0.02$~ & ~$1.39\pm 0.03$~ & ~$0.92\pm 0.02$~  \\
$m_{\tilde l}-m_{\tilde\chi^0_1}$
                       & ~$46.5\pm 12.7$~  & ~$52.7\pm 21.9$~ & ~$69.0\pm 53.6$~ & ~$57.2\pm 95.8$~ \\
$\chi^2/{\rm ndf}$     & 1.20         & 1.29             & 2.06           & 0.90            \\
\hline
\end{tabular}
\caption{Mass differences in [GeV] and fit quality $\chi^2/{\rm ndf}$ 
from fits to the di-lepton invariant mass distributions.}
\label{tab:masses}
\end{center}
\end{table}

For both benchmark points A and B, with the larger statistics, the fit
gives a useful bound on the slepton--singlino mass difference. These
can be compared to the nominal values of $m_{\tilde
l_R}-m_{\tilde\chi^0_1}=37.1$~GeV and $30.4$~GeV, for points A and B
respectively. This sensitivity can be understood physically as the
effect the proximity of the slepton pole has on the shape of the
invariant mass distribution. For much larger slepton masses this
sensitivity should go away. For all four points the fit gives an
accurate determination of $\Delta m$. Comparing to the nominal values
given in Section~\ref{sect:scenario}, in particular the result for
benchmark point B indicates that there are potential sources of
significant systematic error, larger than the statistical errors with
10~fb$^{-1}$ of data.

With the information obtainable from a long decay chain involving an
on-shell slepton, see above, the absolute singlino mass can be found
with the same precision as the other two neutralinos involved,
i.e. around $4\%$ for our scenario, meaning that we are dominated by
the errors of the long decay chain. The results on the
slepton--singlino mass difference in Table~\ref{tab:masses} would
indicate, for benchmarks A and B, that the decay (\ref{eq:singlino})
occurs dominantly through a different slepton than the decay
(\ref{eq:slepton}), which one may speculate could in turn give some
restriction on the neutralino mixing parameters.

\section{CONCLUSIONS}
\label{sect:conclusions}

We have demonstrated that lowering the requirements on lepton
transverse momentum in the standard search for the SUSY di-lepton edge
may reveal unexpected features, such as the NMSSM in disguise. While
our numerical results are sensitive to the exact lepton efficiencies
and momentum resolutions at low transverse momenta --- to be measured
by the LHC experiments --- the OSOF subtraction procedure ensures that
the background can be estimated from data and that the NMSSM scenario
in question is both discoverable down to very small bino--singlino
mass differences, $\Delta
m=m_{\tilde\chi^0_2}-m_{\tilde\chi^0_1}\simeq 0.8$~GeV, and that this
mass difference is measurable to good precision. We have also shown
that the di-lepton invariant mass distribution has some sensitivity to
the slepton--singlino mass difference.

We would also like to note that, since virtually all SUSY cascades in
these scenarios will contain two decays of the type
(\ref{eq:singlino}), this lower edge in the di-lepton distribution may
appear much earlier than the `standard' decay through a slepton, if at
all present, provided that the soft leptons are searched for. This may
in fact be an early discovery channel for SUSY.

\section*{ACKNOWLEDGEMENTS}

We thank members of the ATLAS Collaboration for helpful
discussions. 
%We have made use of the ATLAS physics analysis framework
%and tools which are the results of collaboration-wide efforts.
ARR and MJW acknowledge funding from the UK Science and Technology
Facilities Council (STFC). This work is also part of the French ANR
project ToolsDMColl, BLAN07-2-194882.

\bibliography{KRW}

\end{document}